\documentclass[journal = apchd5, manuscript=letter, layout = onecolumn]{achemso} 

\usepackage{xcolor}
\title{Engineering chiral light-matter interactions in a waveguide-coupled nanocavity} 
\begin{tocentry}
\includegraphics[width = 8.2cm]{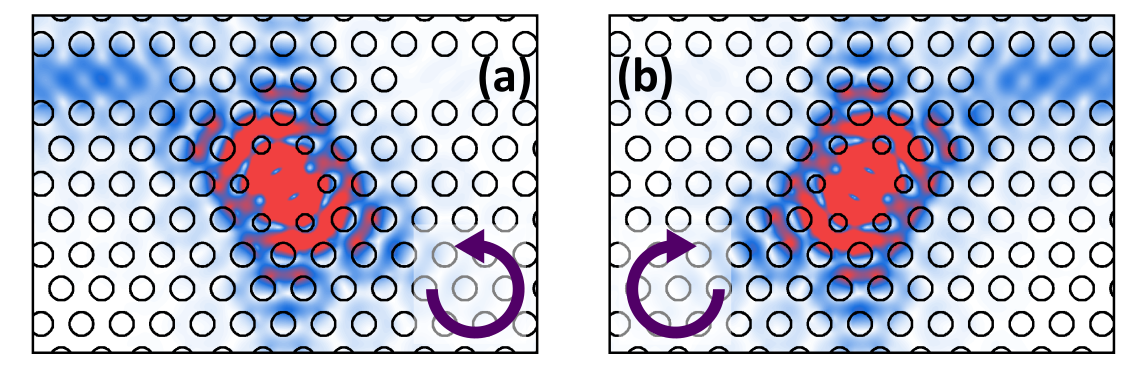}
Directional emission from (a) $\sigma-$ and (b) $\sigma+$ circularly polarised dipoles in an H1 nanocavity.
\end{tocentry}
\author{Dominic Hallett}
\email{d.hallett@sheffield.ac.uk}
\affiliation{Department of Physics and Astronomy, University of Sheffield, Sheffield S3 7RH, UK}
\author{Andrew P. Foster}
\affiliation{Department of Physics and Astronomy, University of Sheffield, Sheffield S3 7RH, UK}
\author{David Whittaker}
\affiliation{Department of Physics and Astronomy, University of Sheffield, Sheffield S3 7RH, UK}
\author{Maurice S. Skolnick}
\affiliation{Department of Physics and Astronomy, University of Sheffield, Sheffield S3 7RH, UK}
\author{Luke R. Wilson}
\email{luke.wilson@sheffield.ac.uk}
\affiliation{Department of Physics and Astronomy, University of Sheffield, Sheffield S3 7RH, UK}
\date{\today}
\begin{document}
\begin{abstract}
Spin-dependent, directional light-matter interactions form the basis of chiral quantum networks. In the solid state, quantum emitters commonly possess circularly polarised optical transitions with spin-dependent handedness. We demonstrate numerically that spin-dependent chiral coupling can be realised by embedding such an emitter in a waveguide-coupled nanocavity, which supports two near-degenerate, orthogonally-polarised cavity modes. The chiral behaviour arises due to direction-dependent interference between the cavity modes upon coupling to two single-mode output waveguides. Notably, an experimentally realistic cavity design simultaneously supports near-unity chiral contrast, efficient ($>95\%$) cavity-waveguide coupling and enhanced light-matter interaction strength (Purcell factor $F_P > 70$). In combination, these parameters enable the development of highly coherent spin-photon interfaces ready for integration into nanophotonic circuits.
\end{abstract}
\maketitle
\section{Introduction}
The ability to control light-matter interactions at the single-photon level is central to the appeal of integrated nanophotonic devices. For instance, an efficient bi-directional light-matter interface can be created by embedding an integrated quantum emitter (QE), such as a quantum dot or diamond colour centre, at the centre of a single mode nanophotonic waveguide \cite{PhysRevLett.113.093603, SiV4}. Recently, it was realised that the strong transverse light confinement in such a waveguide can result in the presence of a locally circularly polarised electric field with a direction-dependent handedness \cite{Chiral1_2017}. For a QE with circularly polarised spin-dependent transitions \cite{Faraday1}, coupling to the waveguide then becomes non-reciprocal. In the ideal case spin and momentum are locked, and the light-matter interface is unidirectional (chiral) for a given QE transition. The chiral interface has many potential applications, from spin-path entanglement \cite{NB1_2016} to path-based spin initialisation \cite{NB2_2017} and the generation of entanglement between multiple emitters \cite{Chiral1_2017, Buonaiuto_2019, PhysRevB.96.115162}. \newline
Chiral coupling has been experimentally demonstrated in several nanophotonic waveguide structures, including dielectric nanobeams \cite{CrossNB2_2013, NB1_2016} and W1 \cite{PhC3_2015}, glide-plane \cite{GP1_2015} and topological \cite{Topological1_2018, Topological2_2020, Topological4_2020} photonic crystal waveguides. A significant remaining challenge is to increase the strength of the light-matter interaction whilst retaining its chiral nature. The benefits of a Purcell-enhancement of the QE-waveguide coupling rate include an increase in the coupling efficiency \cite{PhysRevLett.113.093603} (beta factor) and improvement in the indistinguishability of emitted photons \cite{Purcell1, Purcell2, H1Cav2_2018}. Theoretical work has shown that modest Purcell factors ($5 < F_P < 10$) can be achieved whilst maintaining strong chirality by harnessing slow-light in modified glide-plane photonic crystal waveguides \cite{GP2_2017}. \newline
A competing approach is to position an emitter at a field antinode of an optical cavity or resonator. Chiral coupling has previously been realised between cold atoms and whispering-gallery mode (WGM) resonators, enabling the demonstration of an optical isolator\cite{RR1_2013} and the routing of single photons\cite{RR4_2014}. In the solid state, WGM resonators based on topological photonic crystals\cite{Topological2_2020, Topological3_2020, Topological4_2020} and nanobeam waveguides\cite{RR3_2019} also show significant potential in this regard. In contrast, photonic crystal cavities (PhCCs), which simultaneously support high intrinsic Q factors and low mode volumes, would allow the device footprint to be significantly reduced. Experimentally, PhCCs have been used to achieve Purcell factors as large as 70 in the weak coupling regime\cite{NiV1, H1Cav2_2018, SiV1}, whilst the strong coupling regime has been realised in a number of structures including L3\cite{PhotonBlockade1, PhotonBlockade2} and double heterostructure cavities\cite{PhCHeterostructure1}. A particularly attractive feature of PhCCs for integrated photonics is the ability to efficiently couple the cavity to waveguides, and hence construct extended photonic circuits\cite{Coles:14, H1Cav1_2015}. However, no proposals currently exist for chiral coupling of a QE within a waveguide-coupled PhCC. \newline
Here, we demonstrate through numerical simulations that chiral coupling can be achieved when a QE is embedded inside a PhCC. The proposed device consists of an H1 PhCC coupled to two output waveguides, the presence of which slightly lifts the degeneracy of the two orthogonal cavity modes. 
We show that the relative phase difference in the superposition of the two modes excited by a circularly polarised emitter, combined with the spatial symmetry properties of the modes, leads to interference which is constructive in one waveguide and destructive in the other, and therefore allows for chiral emission from the QE.
Finite-difference time-domain (FDTD) simulations show that our structure supports near-unity chiral contrast, with a simulated Purcell factor as large as 70 for an emitter placed at the centre of the cavity. In addition, we show that the chiral contrast is robust against displacement of the emitter from the cavity centre. Our work provides a path to highly coherent spin-photon interfaces using embedded QEs, and strongly-coupled chiral nanophotonic devices. \newline
\section{Device Design}
\begin{figure*}[t]
\centering
\includegraphics[width=1\columnwidth]{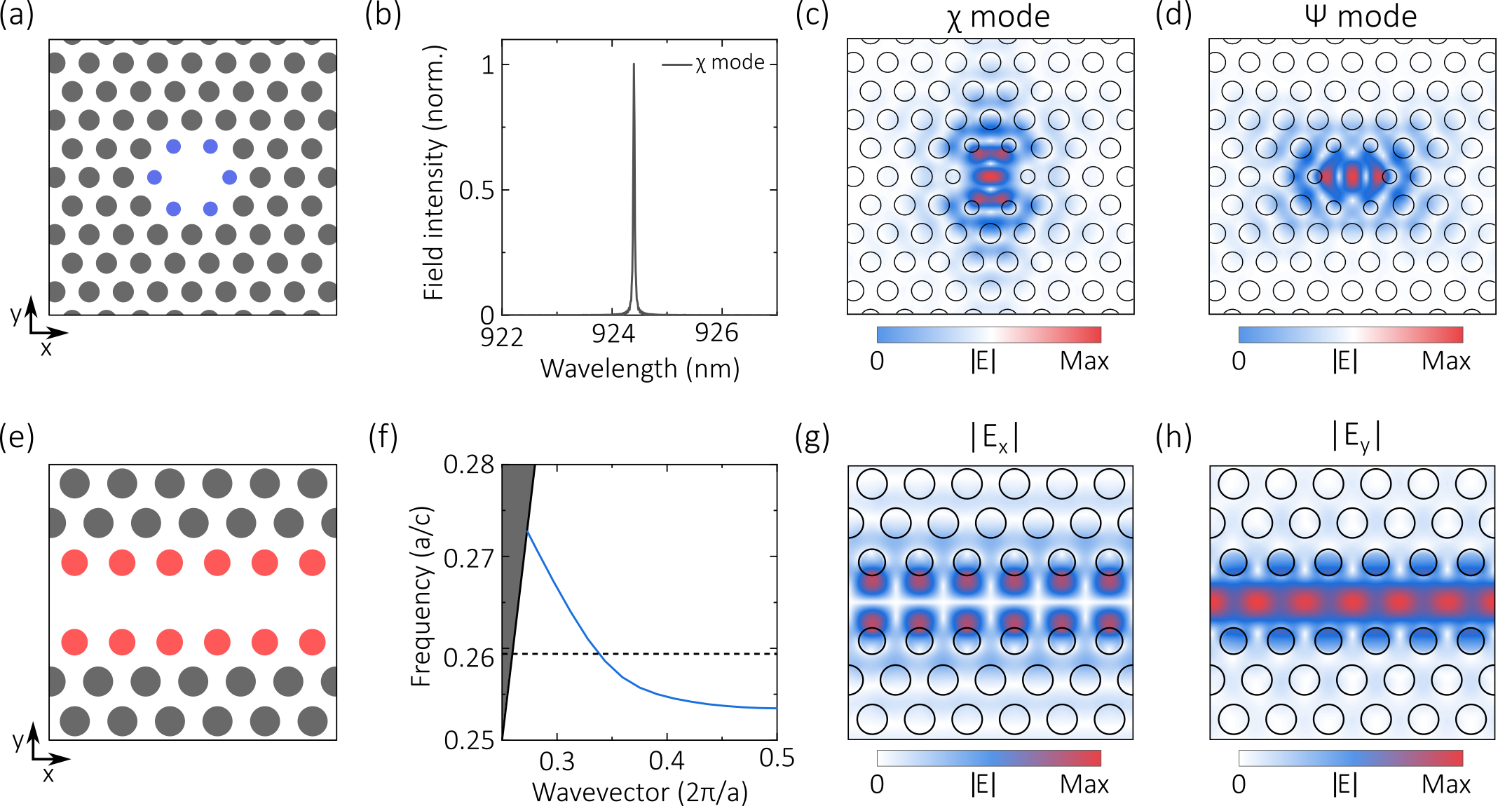}
\caption{\label{fig:One_a}(a-d) H1 cavity design. (a) Schematic of the H1 cavity. Filled circles represent air holes drilled in a dielectric membrane. Blue circles represent the inner holes which are displaced outwards and shrunken compared to the other holes (b) Mode spectrum for the H1 cavity. (c-d) Electric field intensity of the (c) $\chi$ and (d) $\psi$ cavity modes. (e-h) W1 waveguide design (e) Schematic of the W1 waveguide. The red holes are shrunken relative to the other holes. (f) Band structure for the waveguide in (e), where $a$ is the lattice period and $c$ is the speed of light. The mode of interest is the lowest frequency guided mode which is indicated by the solid blue line. The wavelength of the cavity modes is indicated by the dotted black line, and the light line by the solid black line. (g) $|E_x|$ and (h) $|E_y|$ electric field components of the guided mode. }
\end{figure*}
The device is based on a photonic crystal formed from a triangular lattice of air holes within a thin dielectric membrane. The lattice has a period of $a = 240$ nm and hole radius $r = 0.3a$, while the membrane has a thickness $d = 0.71a$ and refractive index $n = 3.4$. These parameters result in a bandgap for TE polarised light covering the range $730 – 1050$ nm and are chosen for compatibility with high-quality QEs operating in the near infrared, such as InAs semiconductor quantum dots. Our chiral device comprises two photonic building blocks. First, an H1 PhCC is formed within the crystal by omitting a single air hole (see Figure~\ref{fig:One_a}a). The PhCC was designed using FDTD simulations (Lumerical FDTD Solutions\cite{Lumerical}). To increase the intrinsic quality factor (Q factor) of the cavity, the innermost holes (marked blue) are displaced outwards by 0.09a and have their radius reduced by 0.09a. The resulting cavity supports two orthogonal fundamental modes labelled $\chi$ and $\psi$ (as in Thijssen et al.\cite{Fields1_2012}), each with a Q factor of $\sim34000$ and centred at $\sim924.5$ nm. The mode spectrum and spatial field profiles of the H1 cavity are shown in Figure~\ref{fig:One_a}b-d. The second element of our device is the W1 waveguide, formed by omitting a single row of holes from the crystal (see Figure~\ref{fig:One_a}e). By shrinking the holes adjacent to the waveguide (marked in red in Figure \ref{fig:One_a}e), we ensure that the waveguide supports a single TE mode at a wavelength corresponding to the fundamental cavity resonance (Fig \ref{fig:One_a}f). The electric field components of this mode are shown in Figure \ref{fig:One_a}g-h. \newline
\begin{figure*}[t]
\centering
\includegraphics[width=1\columnwidth]{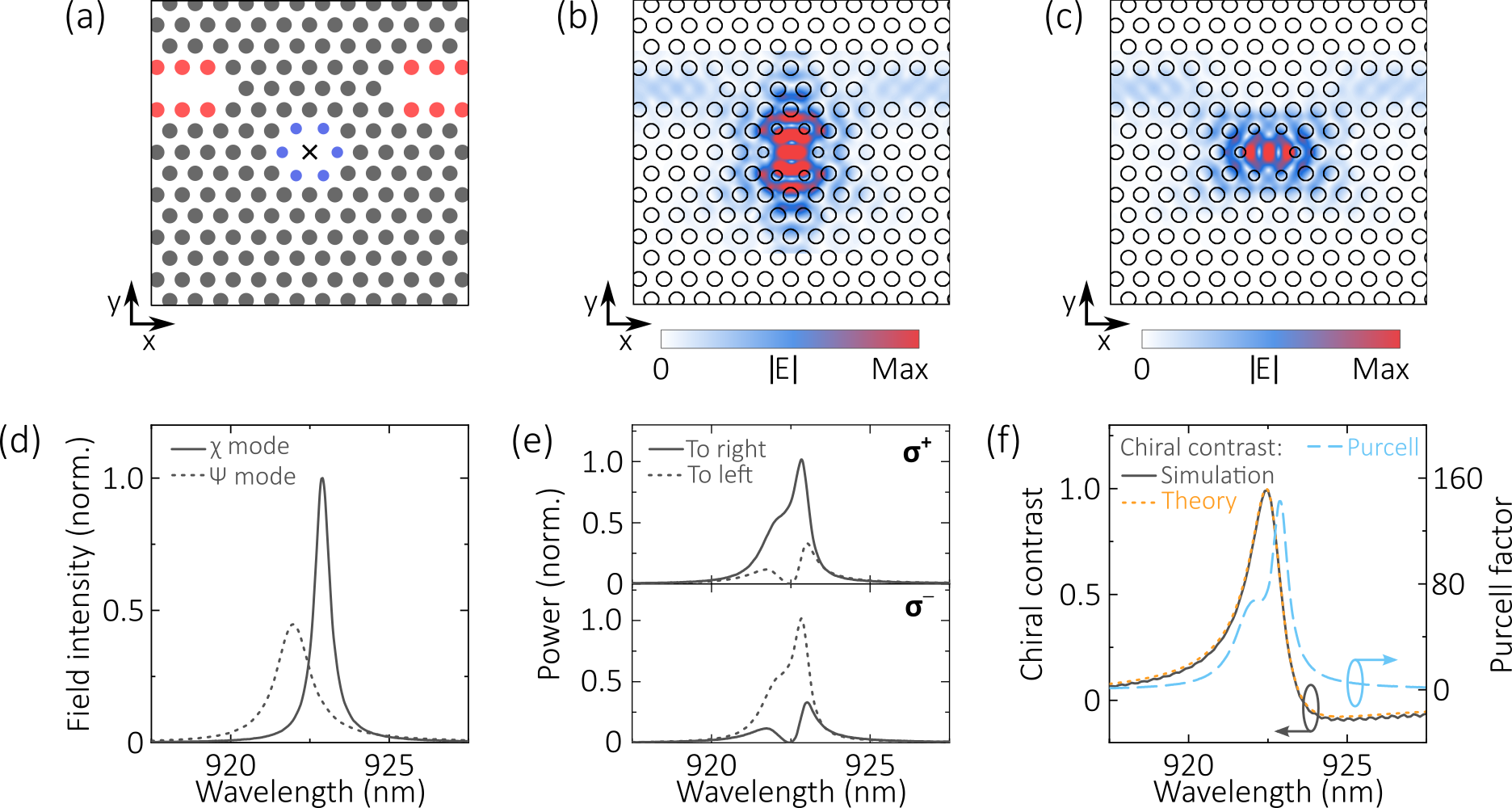}
\caption{\label{fig:One}(a) Schematic of the chiral cavity device. The inner holes of the H1 cavity are marked in blue, and the edge holes of the W1 waveguides are marked in red. The dipole position is marked by a cross. (b,c) Spatial profile of the electric field intensity for the (b) $\chi$ and (c) $\psi$ cavity mode. (d) Spectral profile of the electric field intensity of the cavity modes, normalised to the maximum intensity of the $\chi$ mode. (e) Power emitted by $\sigma^+$ or $\sigma^-$ circular dipoles located at the cavity centre into either the left or right waveguide. (f) Directionality of emission (solid black line - numerical simulation, dotted orange line - theory, left axis) and Purcell factor of a $\sigma^+$ dipole (dashed blue line, right axis) as a function of wavelength. The directionality is evaluated using the data in (e). }
\end{figure*}
To create our chiral device, we bring two W1 waveguides into proximity with the PhCC, as shown in Figure~\ref{fig:One}a. In contrast to previous work\cite{H1Cav1_2015,H1Cav2_2018,Coles:14} the waveguide orientation is chosen such that both cavity modes couple to both waveguides (see Figure~\ref{fig:One}b-c). Introducing the waveguides lifts the degeneracy of the cavity modes. The resulting $\chi$ ($\psi$) mode resonance is centred at $922.9(921.9)$ nm, and has a Q factor of $1600(700)$. The mode spectra are shown in Figure~\ref{fig:One}d. As discussed later in the paper, control of the detuning between the cavity modes is vital to enable chiral coupling in this device. In these simulations, the detuning between the two modes is optimised by uniaxial stretching of the photonic crystal in the y-direction; the separation between rows in the PhC is increased by 1 nm. We note that experimentally, fine control of the mode detuning in H1 cavities has been previously demonstrated using this stretched lattice method \cite{PhCTuning}, as well as through in-situ methods such as strain tuning\cite{StrainTuning} and local oxidation \cite{LAOTuning}. \newline
To demonstrate chiral coupling in our device, we simulate a circularly polarised, broadband dipole positioned at the centre of the cavity and monitor the optical power coupled into each waveguide. The power is calculated from the Poynting vector of the fields measured at the output of each waveguide. Figure~\ref{fig:One}e shows the coupled power for either a $\sigma^+$ or $\sigma^-$ polarised dipole. Looking first at the $\sigma^+$ dipole, two significant features can be observed. Firstly, the power emitted into the right waveguide is greater than that coupled into the left waveguide over a bandwidth of a few nanometers, indicating that the emission exhibits a degree of chirality over this bandwidth. Secondly, at a wavelength of $922.45$ nm the device is almost perfectly chiral – nearly all of the emission is coupled into the right-hand waveguide, whilst emission into the left-hand waveguide is strongly suppressed. The result for the $\sigma^-$ dipole is reversed, but otherwise identical. To quantify the directionality of the emission, we calculate the chiral contrast using Equation~\ref{eq:Directionality}: \newline
\begin{equation}
C = \frac{P_R-P_L}{P_R+P_L}
\label{eq:Directionality},
\end{equation}
where $P_L$ and $P_R$ are the powers coupled into the left and right hand waveguides, respectively. \newline
The wavelength dependence of the contrast is shown in Figure~\ref{fig:One}f. We note that the maximum chiral contrast (C = 0.996 at 922.45 nm) occurs when the power emitted by the dipole into the $\chi$ and $\psi$ modes is approximately equal (compare with Figure~\ref{fig:One}d).  Also shown in Figure~\ref{fig:One}f is the wavelength-dependent Purcell factor, which has a value of $F_P \approx 72$ at 922.45 nm. This was obtained by normalising the total power output of a circularly polarised dipole (measured using a ‘box’ of monitors surrounding the source) to the power emitted by such a dipole in a homogenous medium. Notably, the chiral contrast is greater than 0.9 over a bandwidth of 0.4 nm (922.23 nm – 922.63 nm), while the Purcell factor is larger than 65 over the same bandwidth. \newline
By comparing the power output of a circularly polarised dipole to the total power coupled into the waveguides, we calculate the emitter-cavity-waveguide coupling efficiency to be 95\%. There exists a a trade-off between the achievable coupling efficiency and the coupled cavity Q factors, but the efficiency could be increased by further optimisation of the cavity design\cite{H1_HighQ, H1Cav3_2012}. Finally, we find that the waveguide-to-waveguide transmission is T=95\% at 922.45 nm (see Appendix 1 for spectral dependence). We conclude that the device simultaneously supports near-unity chiral contrast and strong light-matter interactions.
\section{Origin of chiral coupling}
\begin{figure}
\centering
\includegraphics[width=0.7\columnwidth]{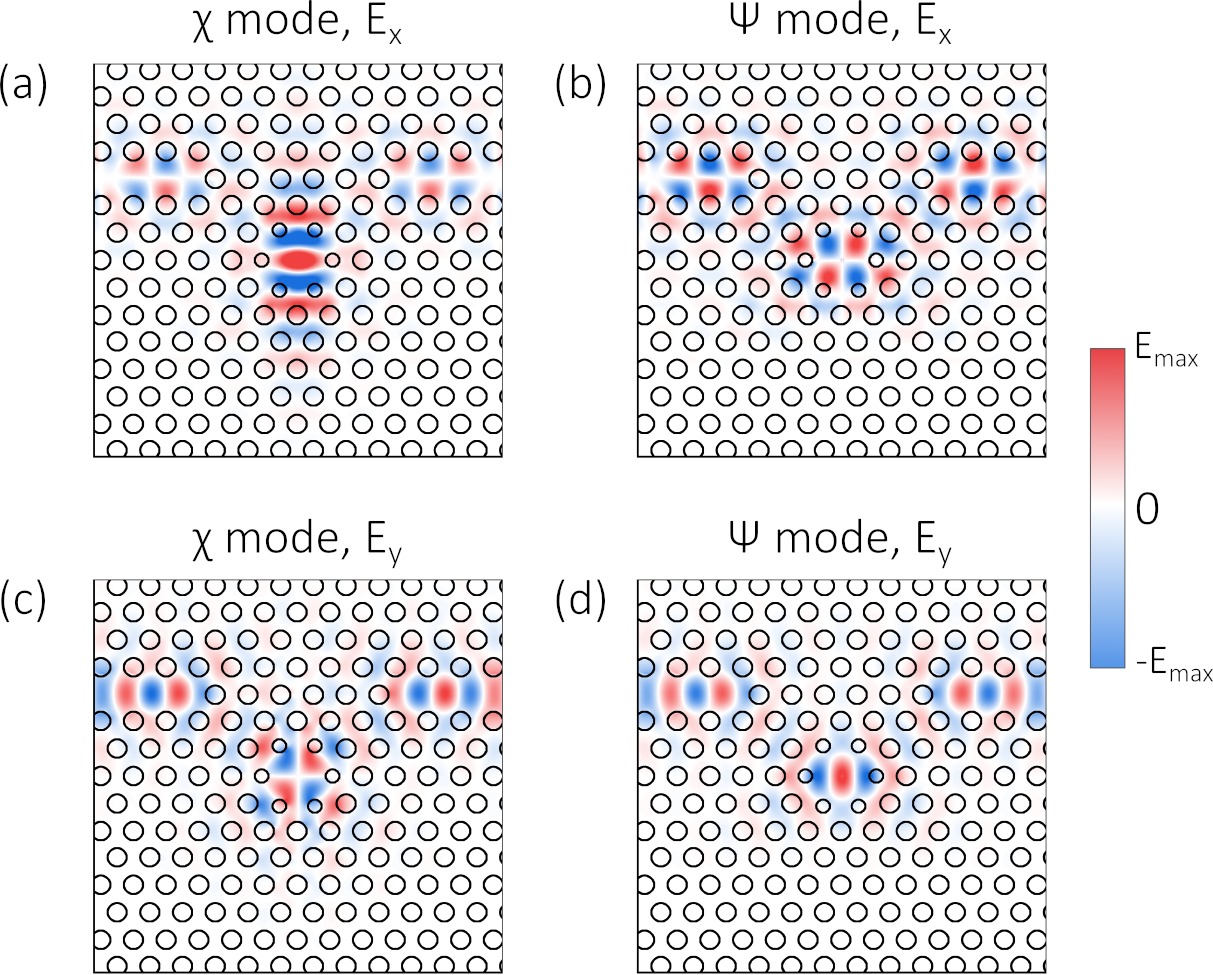}
\caption{\label{fig:Two}Spatial dependence of the cavity mode electric field components. (a) $E_x^{\chi}$  (b) $E_x^{\psi}$ (c) $E_y^{\chi}$ (d) $E_y^{\psi}$. The black circles are outlines of the PhC holes.}
\end{figure}
The mechanism behind the predicted chiral coupling can be understood by considering the spatial symmetry of the cavity modes, their phase relationship and their relative intensity. We first consider the electric field spatial profiles of the two cavity modes. The field profiles are obtained separately for the $\chi$ and $\psi$ modes using independent simulations. This is possible because  an x polarised dipole positioned at the cavity centre excites only the $\chi$ mode, while a y polarised dipole at the cavity centre excites only the $\psi$ mode. Figure~\ref{fig:Two} shows spatial maps for the $E_i^j(x,y)$ ($i=x,y; j= \chi,\psi$) components of the electric field for both cavity modes in the plane of the device. At the cavity centre, the modes are completely orthogonally polarised, supporting only a single field component ($E_x$ or $E_y$). However, away from the cavity centre, the modes have contributions from both $E_x$ and $E_y$ field components. Note also that within the waveguides, the fields take on the profile of the fundamental W1 waveguide mode. This results in perfect spatial overlap between the field components in the waveguides, independent of the cavity mode responsible for the waveguide excitation. When both cavity modes are excited, interference will therefore occur between the fields coupled into the waveguides.  \newline
What form the interference takes in either waveguide depends on the symmetry properties of the cavity field components. For the $E_x$ components, we see that the $\chi$ mode (Figure ~\ref{fig:Two}a) is symmetric in x, whilst the $\psi$ mode (Figure ~\ref{fig:Two}b) is antisymmetric, with this symmetry extending to the fields coupled into the waveguides. A similar observation can be made for the $E_y$ fields, for which the $\chi$ ($\psi$) mode is antisymmetric (symmetric) in x.
Consequently, the fields from the $\chi$ and $\psi$ modes in the left waveguide are in phase, while the fields in the right waveguide are in antiphase. This indicates that interference between the fields of the two modes can be different in the two waveguides, for example destructive in one waveguide and constructive in the other. \newline
Next, we consider the nature of the cavity mode excitation itself. The results discussed above were obtained from independent excitation of each mode, whilst we wish to excite the modes simultaneously using a single QE with a circularly polarised transition. The circular dipole will excite a superposition of the two cavity modes \cite{Fields1_2012} with a known phase difference ($\pm\pi/2$, depending on the dipole handedness). This alone will not lead to directional emission, as the phase difference between the fields in one waveguide will be $+\pi/2$, and in the other will be $-\pi/2$. However, an additional phase difference can arise, from the detuning between the emitter and each cavity mode. Critically, this requires that the modes be non-degenerate, as we show below. An additional phase difference of $\pm\pi/2$ will mean the fields of the two modes are in phase in one waveguide, and in antiphase in the other waveguide. Then, if the amplitudes of the fields are the same, complete destructive interference can be achieved in one waveguide, and directional emission is realised. We note that the conditions for directional emission in this system (orthogonally polarised components of equal amplitude with a $\pi/2$ phase difference) describe a circularly polarised field, which is often used to predict directional emission in waveguides \cite{Chiral1_2017, NB1_2016, PhC3_2015, GP1_2015, Topological1_2018, Topological2_2020}. \newline

\section{Analytical Model for Directionality}
\begin{figure*}[ht!]
\centering
\includegraphics[width=1\columnwidth]{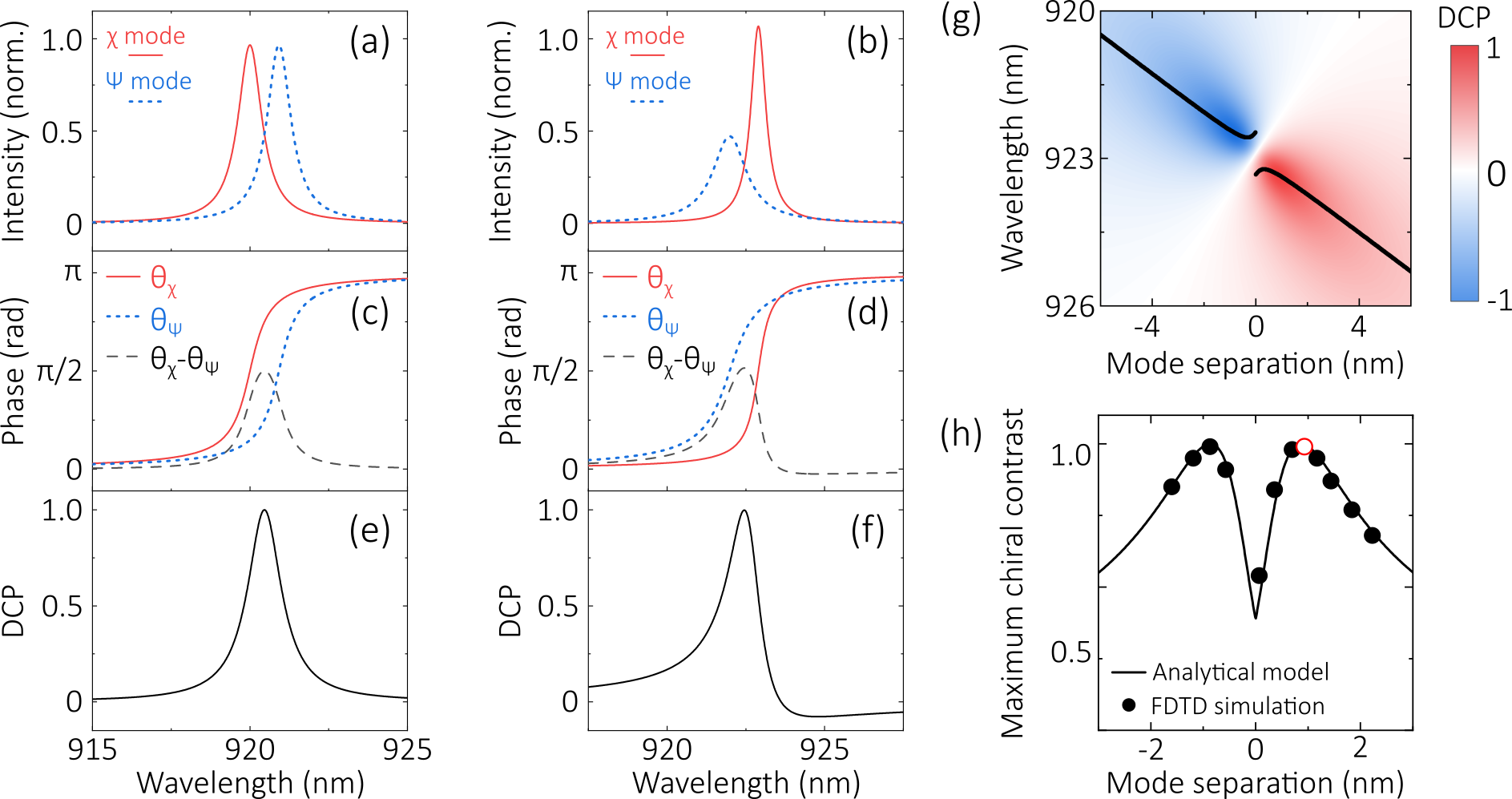}
\caption{\label{fig:Three} (a) Modelled electric field intensity for two near-degenerate modes with $\lambda_{\chi} = 920.0$ nm, $\lambda_{\psi} = 920.92$ nm and $Q_{\chi} = Q_{\psi} = 1000$. (b) Modelled electric field intensity for the cavity modes presented in Figure 2. ($\lambda_{\chi} = 922.9$ nm, $\lambda_{\psi} = 921.9$ nm, $Q_{\chi} = 1600$, $Q_{\psi} = 700$) (c-d) Phase associated with each cavity mode, and resulting phase difference, for the (c) Q = 1000 and (d) simulated modes. (e-f) Degree of circular polarisation (DCP) arising from the detuning between the (e) Q = 1000 and (f) simulated modes. (g) DCP as a function of detuning between the two simulated modes shown in (b). The point of maximum circular polarisation is marked at each mode separation (black line). 
(h) Comparison of analytical results to results from FDTD simulations. The black line indicates the maximum DCP predicted by the analytical model, as a function of the separation between the two cavity modes. The points mark the maximum chiral contrast obtained from FDTD simulations. For these simulations, the lattice constant of the photonic crystal is varied to produce devices with different mode separations. The result of the simulation used in Figure 2 is marked with a red circle. } 
\end{figure*}
In the following section a simple model is developed to describe the effect of detuning the two cavity modes. The model accounts for the detuning-dependent phase difference between the cavity modes, as well as taking into account differences in mode intensity. The model predicts the key behaviour of our device, closely matching the simulated results presented in Figure~\ref{fig:One}f, and provides a straightforward means to optimise the device design. In the model, the electric field of each cavity mode is described by a complex Lorentzian function. It is assumed that the two modes are orthogonally polarised such that the x and y components of a circular dipole couple exclusively to the $\chi$ and $\psi$ modes respectively, which is true for a dipole at the centre of the cavity. This corresponds to the ideal spatial position of the dipole to achieve the maximum Purcell factor. The Lorentzian function describing the amplitude of each mode can be written as
\begin{equation}
E_j(\lambda) = \frac{\sqrt{\Gamma_j}}{\Gamma_j/2 + i(\lambda-\lambda_0^j)}
\label{eq:Lorentzian},
\end{equation}

where $\lambda_0^j$ is the resonant wavelength and $\Gamma_j$= $\lambda_0^j$/Q the decay rate of mode $j=\chi,\psi$.
The mode profiles are shown in Figure~\ref{fig:Three}a for a simplified case in which the two modes have equal Q factors, and the mode separation is chosen to be equal to $\Gamma_\chi$ ($\cong\Gamma_\psi$). \newline
The wavelength dependent phase $\theta_j$ associated with each mode is given by
\begin{equation}
\theta_j = tan^{-1}\left(\frac{\Gamma_j\lambda}{ \lambda^2-{\lambda_0^j}^2}\right)
\label{eq:Phase},
\end{equation}
The different resonant wavelengths of the two modes result in a wavelength-dependent phase difference $\Delta\theta = \theta_\chi - \theta_\psi$, which is shown in Figure~\ref{fig:Three}c. Note that at a wavelength of $(\lambda_\chi + \lambda_\psi)/2$, the modes have equal amplitude and a phase difference of $\pi/2$. A QE emitting at this wavelength would therefore exhibit perfect chiral coupling. \newline
To account both for the phase difference and the difference in amplitude of the two modes as a function of wavelength, the normalised degree of circular polarisation (DCP) arising purely from the cavity modes is calculated using 
\begin{equation}
DCP = \frac{S_{(3)}}{S_{(0)}} = \frac{-2Im(E_{\chi}E_{\psi}^*)}{|E_\chi|^2+|E_\psi|^2}
\label{eq:S3},
\end{equation}
where $S_{(0)}$ and $S_{(3)}$ are the Stokes parameters for intensity and degree of circular polarisation, respectively. The degree of circularity is shown in Figure~\ref{fig:Three}e. With the additional $\pi/2$ phase difference arising from use of a circular dipole to excite the cavity modes, DCP is equivalent to the chiral contrast of the device.\newline
We now compare the results of the model with the FDTD simulated device considered in Figure~\ref{fig:One}. In Figure~\ref{fig:Three}b,d,f we show the modelled electric field profiles, their phase difference and the resulting DCP. The electric field intensity of the two modes is equal at a wavelength of 922.54 nm. The maximum phase difference is $0.515\pi$ radians, obtained at 922.45 nm. At this wavelength DCP is approximately unity (DCP $>$ 0.998), and is greater than 0.9 over a bandwidth of 0.4 nm (922.23 – 922.63 nm). We compare the model to the chiral contrast obtained from simulation of the full device in Figure~\ref{fig:One}f, and find good agreement between the two approaches.\newline
We can now use our model to estimate the degree of chirality as a function of detuning between the two cavity modes. This is valid as the Q factor of each mode remains almost constant when the detuning is altered by a slight change in the lattice parameters. Figure~\ref{fig:Three}g shows the degree of circularity arising from mode detuning as the $\psi$ mode is tuned between 918 nm and 928 nm, whilst the $\chi$ mode is fixed at 922.9 nm. For zero mode separation, no chirality can be observed for a QE on resonance with both modes. The difference in Q factors does, however, allow a degree of chirality to be realised with the emitter slightly off-resonance. Most significantly, the degree of chirality has a global maximum for non-zero mode detuning. This is investigated further in Figure~\ref{fig:Three}h, which presents the maximum chiral contrast for each mode separation (these points are also marked by the black line in Figure~\ref{fig:Three}g). To achieve highly chiral ($C>0.9$) emission, the model predicts that the mode separation $|\lambda_0^\chi-\lambda_0^\psi|$ needs to be between 0.5 nm and 1.4 nm. This level of control of the mode separation is well within current fabrication capabilities \cite{PhCTuning}. \newline
The results of the model are also compared in Figure~\ref{fig:Three}h to simulations in which perturbation of the lattice constant of the photonic crystal is used to tune the mode separation. Close agreement is found between the circularity deduced from the model and the chiral contrast obtained from full simulations, providing strong support for the validity of the model. For a specific location of the cavity-adjacent W1 waveguides, only two simulations are therefore required to obtain the Q factors and wavelengths of the two H1 dipole modes. The model can then be used to deduce the mode separation required in order to achieve maximum chiral contrast without recourse to further (time consuming) simulations.
\section{Effect of Emitter Position}
\begin{figure}[h]
\centering
\includegraphics[width=1\columnwidth]{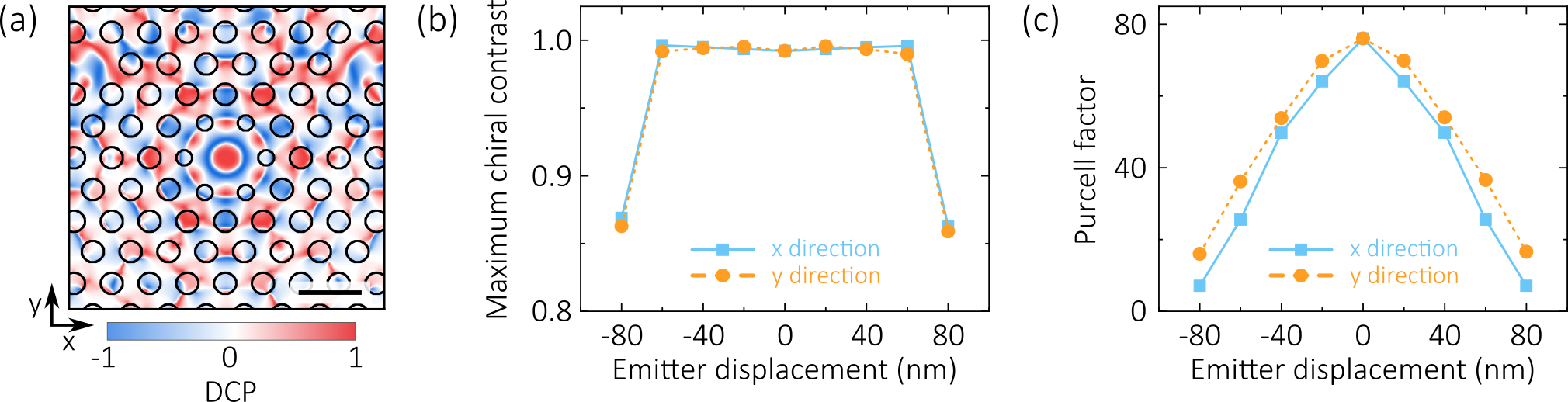}
\caption{\label{fig:Five} (a) Position dependence of the DCP at 922.45 nm. The black circles are outlines of the PhC holes. Scale bar 400 nm. (b-c) Simulated maximum chiral contrast (b) and Purcell factor (c) as a function of emitter displacement from the cavity centre.}
\end{figure}
Finally, we demonstrate that the chiral contrast in our device is robust against position variation of the QE, which can be a limitation in chiral photonic devices \cite{NB1_2016, GP1_2015, PhC1_2015}. The position dependence is initially evaluated by injecting light into the device through one waveguide, then monitoring the spatial dependence of the resulting $E_x$ and $E_y$ cavity fields at the wavelength corresponding to maximum chiral contrast. The degree of circular polarisation is then determined from  Equation~\ref{eq:S3}. \newline
Figure~\ref{fig:Five}a shows the position dependence of the DCP at 922.45 nm for the device considered in Figure~\ref{fig:One}. A circular region of $\sim80$ nm radius in which the DCP is high can be seen in the centre of the cavity. Notably, this coincides with the region of highest field intensity in the cavity, therefore enabling high Purcell factors and near-unity chiral contrast to be achieved simultaneously. \newline
The position dependence is examined in more detail by simulating the properties of emitters at different locations within the cavity. The maximum chiral contrast and Purcell factor obtained from these simulations are shown in Figure~\ref{fig:Five}b and \ref{fig:Five}c. Note that a small position-dependent change in the QE wavelength (of no more than 0.2 nm) is required to achieve the maximum chiral contrast shown here. The chiral contrast can be seen to be robust against significant displacements of the emitter within the cavity; remarkably, even with the emitter positioned up to 60 nm from the centre of the cavity the chiral contrast remains above 0.99. For comparison, a displacement as little as 20 nm from the ideal chiral point in a nanobeam waveguide or a glide-plane photonic crystal may be sufficient to reduce chiral contrast below 0.9 (see Appendix 2). With an emitter offset by 60 nm from the cavity centre, our device still has chiral contrast $>0.99$, a Purcell factor $>25$, and coupling efficiency $>95\%$, and is therefore dramatically more robust to an imperfectly positioned emitter than existing chiral waveguide devices. \newline
\section{Conclusion}
In conclusion, we have proposed a waveguide-coupled cavity nanophotonic device which simultaneously supports near-unity chiral coupling and significant Purcell enhancement ($F_P > 70$) for an embedded circularly polarised emitter. The device consists of an H1 PhCC and two access waveguides. Chiral behaviour arises due to the difference in interference between the two orthogonal cavity modes upon coupling to the access waveguides (being either destructive or constructive, respectively). As this new mechanism for the generation of directional light-matter interactions is quite general, it may also prove applicable to other cavity designs that support orthogonal modes (e.g. micropillars). \newline
The performance of our device is robust to both spectral and spatial perturbation of the embedded emitter. Significantly, in contrast to alternative approaches \cite{NB1_2016, GP1_2015, GP2_2017, PhC1_2015, PhC2_2017, Topological1_2018, Topological2_2020, Topological3_2020, RR3_2019} the chiral contrast of our device is not dependent on precise positioning of the emitter (remaining above 0.99 for displacements of up to 60 nm from the cavity centre). Furthermore, the chiral contrast depends on the relative wavelengths of the cavity modes and the emitter, parameters which can be controlled in-situ \cite{H1Cav2_2018, StrainTuning}, thus enabling tuning and optimisation of the chiral contrast post-fabrication. \newline
Due to the very low (V $\approx0.6(\lambda/n)^3$) mode volume of the H1 cavity, we achieve a large emitter coupling strength in a low-Q (Q $\approx1000$) cavity. This combination of parameters allows for efficient cavity-waveguide coupling ($>95\%$), resulting in waveguide-to-waveguide transmission greater than $95\%$, while obtaining a Purcell factor significantly larger than that possible in devices that do not use a cavity. Furthermore, the relatively low Q allows $F_P$ to remain above 50, and the chiral contrast above 0.9, over a spectral bandwidth of 0.4 nm; this would not be possible with high Q approaches. A low Q, low mode volume design is also beneficial for achieving high Purcell factors without entering the strong-coupling regime. This work presents a method of producing chiral interfaces that are robust, tuneable and scalable, ready for integration into quantum photonic circuits.
%
\section{Funding Sources}
This work was funded by the Engineering and Physical Sciences Review Council (EP/N031776/1).
\begin{acknowledgement}
The authors thank H. Siampour for providing the simulation data for a glide-plane waveguide structure.
\end{acknowledgement}
\appendix
\section{Appendix 1: Device transmission}
\begin{figure}[h]
\centering
\includegraphics[width = 0.5\textwidth]{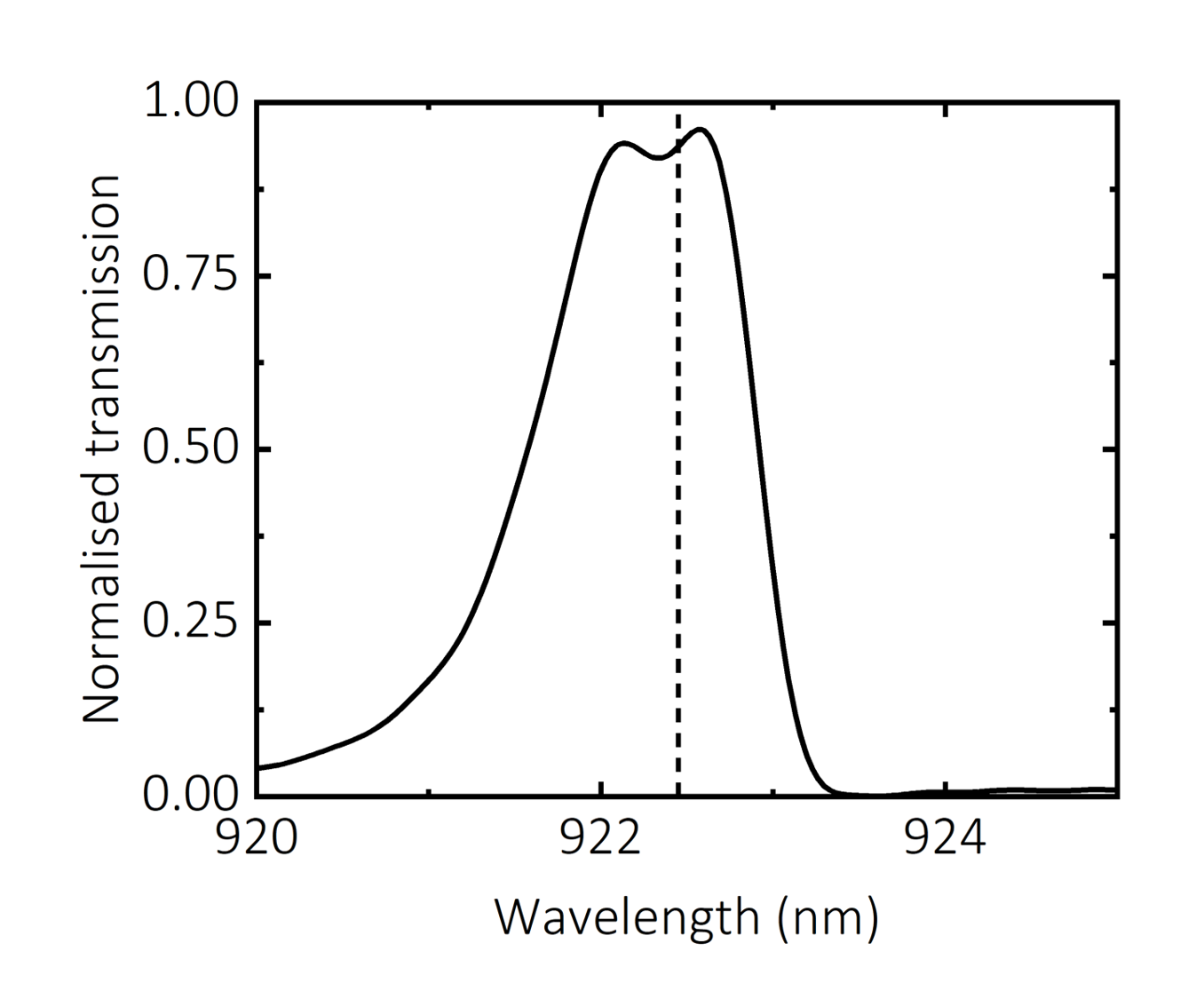}
\caption{Transmission through the device.}
\label{fig:Suppl1}
\end{figure}
The device transmission is calculated using an FDTD simulation in which broadband light is injected into the left-hand waveguide via a mode source, and the power coupled into the right-hand waveguide is monitored. As the mode source does not couple perfectly to the waveguide’s fundamental mode, we also simulate the transmission of an equal length of W1 waveguide as a reference. The mode coupling efficiency is the same in both simulations (as the waveguides are identical) and therefore the only difference between them is the presence or absence of the cavity. The normalised transmission found from these simulations is shown in Figure~\ref{fig:Suppl1}. At the wavelength of interest in our device (922.45 nm), the transmission in the cavity simulation is 0.82, normalised to the total power injected into the simulation domain. In the W1 reference simulation, a value of 0.86 is obtained. The normalised transmission of the chiral device is therefore 0.95 at 922.45 nm. Significantly, the normalised transmission is $>0.9$ over a bandwidth of 0.4 nm; this is consistent with the bandwidth for which the chiral contrast of the device is also $>0.9$.
\section{Appendix 2: Position sensitivity of the chiral contrast in nanobeam and glide-plane waveguides}
\begin{figure}[h]
\centering
\includegraphics[width = 0.3\textwidth]{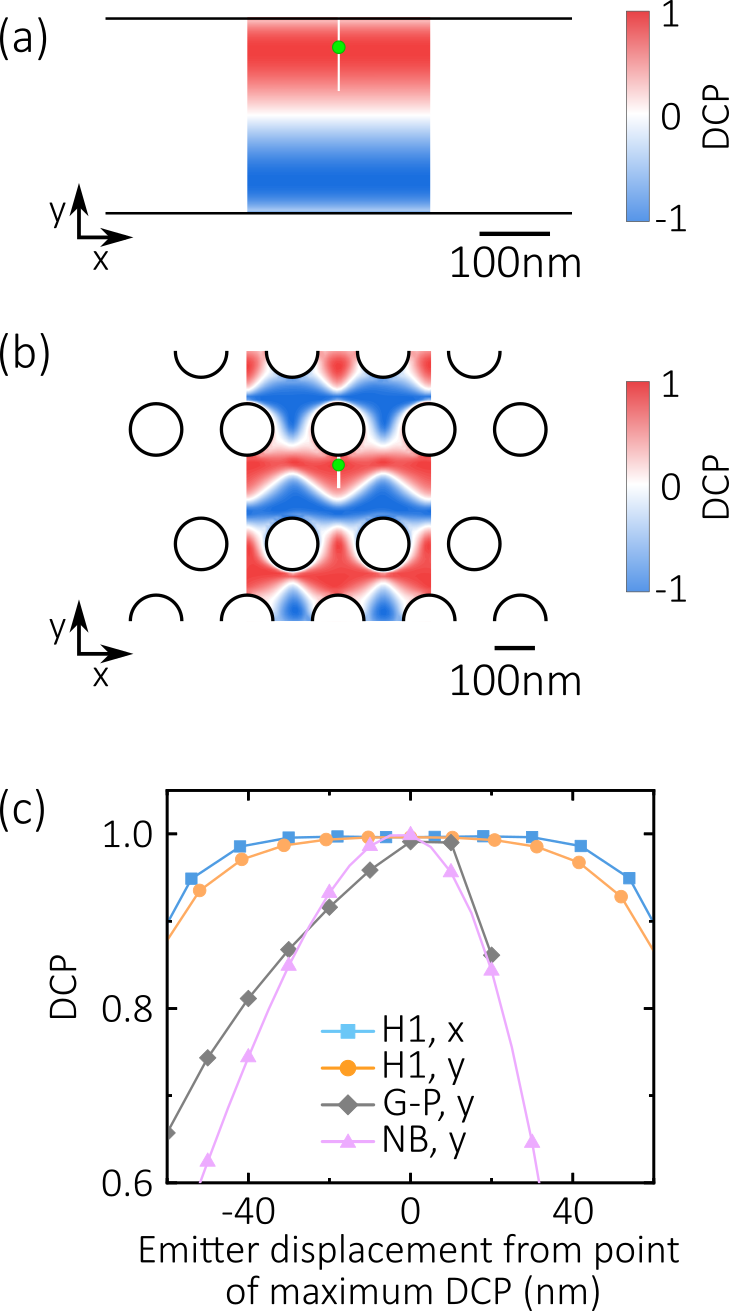}
\caption{(a) DCP at a wavelength of 900 nm in a nanobeam waveguide of 280 nm width and 170 nm thickness. The maximum DCP is indicated by a green circle (note that this is independent of position along the waveguide). (b) DCP at 900 nm in a glide-plane PhC waveguide with period of a=228 nm, hole radius of $0.3a$ and membrane thickness of 170 nm. The maximum DCP is indicated by a green circle. (c) DCP as a function of emitter position relative to the position of maximum DCP, for a nanobeam waveguide, glide-plane photonic crystal waveguide, and our H1 PhCC device. For the waveguide devices, we consider displacement of the emitter in a direction perpendicular to the waveguide, as indicated by the white lines in (a) and (b). }
\label{fig:Suppl2}
\end{figure}
To compare the position sensitivity of our device with that of competing nanophotonic structures, we also evaluate the position-dependent DCP for nanobeam and glide-plane waveguides. The simulated nanobeam has a rectangular cross-section with a width (height) of 280 nm (170 nm), supporting single mode operation at our operating wavelength (around 900 nm). For the glide-plane waveguide, we use a lattice period of a=228 nm, radius of $0.3a$ and membrane thickness of 170 nm. These glide-plane parameters are optimised for single-mode slow-light operation at 900 nm. \newline
For the nanobeam, the position of maximum DCP is approximately 95 nm from the centre of the waveguide (see circle in Figure 7a). We evaluate the DCP along a line cut in the direction perpendicular to the waveguide, and plot the DCP in Figure 7c. Note that for the ideal nanobeam, there is no position sensitivity along the waveguide. \newline
For the glide-plane, the optimal chiral position is in the vicinity of a PhC hole (see circle in Figure 7b). Again, we evaluate the DCP in a line perpendicular to the waveguide and plot this in Figure 7c. Finally, we also show in Figure 7c the DCP evaluated for the H1 cavity device.  The H1 device is clearly more robust to position variation of an embedded emitter. \newline
The data for the H1 device in this case is taken from the simulation used in Figure 5a, rather than the similar data in Figure 5b, so that the simulation method is consistent when comparing the three devices. A small difference in the result arises because the wavelength of maximum chiral contrast for the H1 device shifts slightly as the emitter is displaced. The method used here considers only a single wavelength (922.45 nm), while in Figure 5b the maximum chiral contrast in each simulation is used.
\bibliography{ACSPhot-Revised}  
\end{document}